\documentclass[aps,pra,reprint,superscriptaddress]{revtex4-2}

\expandafter\let\csname equation*\endcsname=\relax
\expandafter\let\csname endequation*\endcsname=\relax
\usepackage{amsmath}
\usepackage{amssymb}
\usepackage{graphicx}
\usepackage{subfigure}
\usepackage{geometry}
\usepackage{hyperref}
\usepackage{color}
\usepackage{float}

\begin{abstract}
	Guided wave optics, including most prominently fiber optics and integrated photonics, very often considers only one or very few spatial modes of the waveguides. Despite being known and utilized for decades, multi-mode guided wave optics is currently rapidly increasing in parallel with technological improvements and better simulation tools. The physics of multi-mode interactions are usually driven by some initial energy distribution in a number of spatial modes. In this work we introduce how, with free-space input beams having space-time couplings, the different modes can be excited with different complex frequency or time profiles. We cover fundamentals, the coupling with a few simple space-time aberrations, different waveguides, and a number of technical nuances. This concept of space-time initial conditions in multi-mode waveguides will provide yet another tool to study the rich nonlinear interactions in such systems.
\end{abstract}
\begin{document}

\title{Coupling to multi-mode waveguides with space-time shaped free-space pulses}
\author{Spencer W. Jolly}
\email{spencer.jolly@ulb.be}
\affiliation{Service OPERA-Photonique, Université libre de Bruxelles (ULB), Brussels, Belgium}
\author{Pascal Kockaert}
\email{pascal.kockaert@ulb.be}
\affiliation{Service OPERA-Photonique, Université libre de Bruxelles (ULB), Brussels, Belgium}

\date{\today}
\maketitle

\section{Introduction}
\label{sec:intro}

The inclusion of multiple spatial modes in guided-wave optics adds another degree of freedom~\cite{piccardo22,cristiani22} and in the frame of nonlinear and ultrafast optics can enable new technologies~\cite{wright22-1} and a large number of new and interesting phenomena~\cite{krupa19,wright22-2}. Examples include multi-mode solitons and supercontinuum generation~\cite{wright15-1,wright15-2,wright15-3,sunY22}, spatio-temporal modelocking~\cite{wright17,tegin19,wright20} and other novel light sources~\cite{haig22}, spatial beam self-cleaning~\cite{liuZ16,krupa17,tegin20} or compression~\cite{krupa18}, and even measurement, imaging, and computation~\cite{xiong20-1,tegin21,rahmani22}.

On the other hand, in free-space ultrashort optics, space-time couplings~\cite{akturk10} and in general space-time beam shaping~\cite{shenY23} can enable a similarly transformative control over the evolving electric field of the pulses. Space-time optics can therefore influence light-matter interaction physics and beyond~\cite{vincenti12,froula18,sun18}.

Free-space optics and guided-wave optics are justifiably treated most often as separate subjects, and this includes in the frame of space-time optics. However, in the context of coupling from free-space into a multi-mode waveguide, a form of mode conversion, there is an opportunity and necessity to consider both subjects. In this tutorial we will bridge space-time descriptions of free-space and guided optics to describe how a beam with space-time couplings will couple to the different spatial modes of a given waveguide. This will result in initial conditions in a multi-mode waveguide that depend on space and time---i.e. coupling coefficients to each mode that have a complex temporal or spectral amplitude. We will consider only waveguides that have bounded propagating modes, in contrast to recent work that uses advanced space-time shaping to produce propagation invariant supermodes in unbounded planar waveguides~\cite{shiri20,shiri22}.

Despite this work being novel and relatively unexplored, we will expand upon previous work~\cite{guangZ18} and develop the necessary tools and results in a tutorial manner. We will begin with the basic characteristics and tools for free-space beams and for multi-mode waveguides in Section~\ref{sec:fundamentals}, describe the complex coupling with various "low-order" space-time couplings in Section~\ref{sec:STC_coupling}, and finally discuss more nuanced scenarios such as arbitrary space-time inputs, few-cycle pulses, and a larger range of multi-mode waveguide platforms in Section~\ref{sec:discussion}. Importantly, we will consider only the linear problem of coupling to a multi-mode waveguide. The linear propagation after such coupling depends on the absorption, group velocity, and dispersion of each mode. The nonlinear optical physics that would follow when the pulse energy becomes large enough is a much more complex problem, solvable in a number of ways~\cite{kolesik02,poletti08,wright18,github_GMMNLSE,bejot19}, which will be a topic of our future work.

\section{Fundamentals of multi-mode fibers, mode coupling, and space-time effects}
\label{sec:fundamentals}

Before discussing ultrashort pulse-beams with space-time couplings, it is important to introduce the basics of multi-mode waveguides and the coupling of free space beams. For the purposes of this section we will consider only cylindrically symmetric waveguides that have multiple propagating modes due to their larger diameter. In the discussion section we will discuss other forms of multi-mode waveguides.

\subsection{Modes of step-index and GRIN fibers}
\label{sec:modes}

\begin{figure*}[htb]
	\centering
	\includegraphics[width=\textwidth]{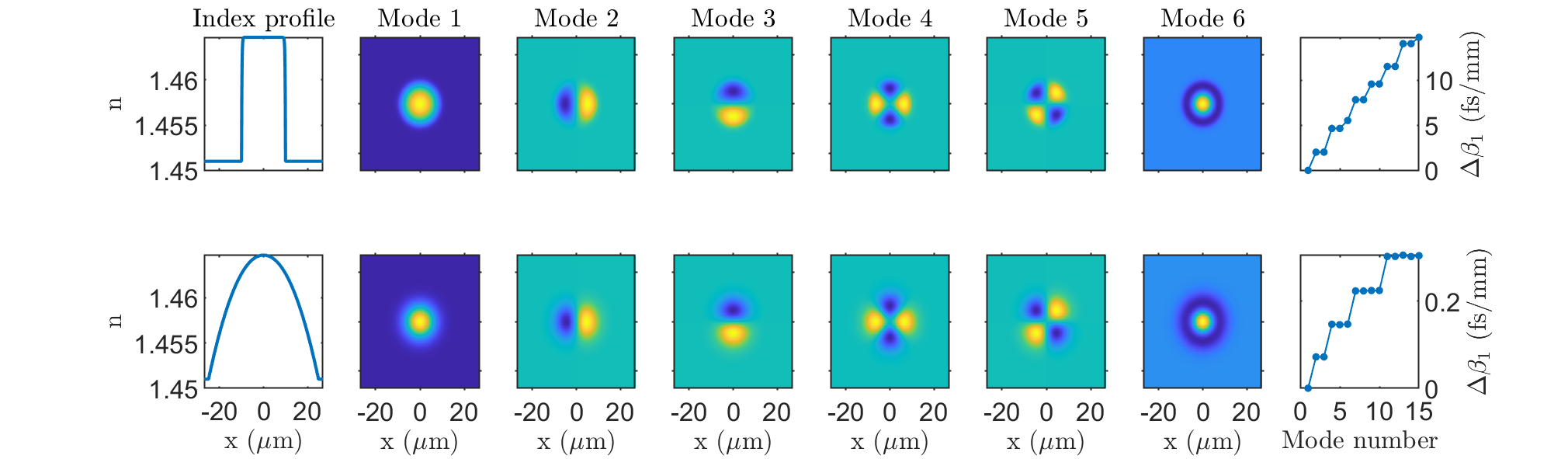}
	\caption{Comparison of step-index (top) and GRIN fibers (bottom) with similar fundamental modes sizes at the central wavelength of 1030\,nm. The index profiles are shown on the far left, followed by the first 6 modes, with finally the first-order propagation parameter relative to that at the central frequency (inversely proportional to the group velocity difference) shown on the right.}
	\label{fig:GRIN_step_compare}
\end{figure*}

The index profiles and scalar modes, i.e. the modes at only one polarization, of a step-index fiber of 10\,$\mu$m radius and a parabolic graded-index (GRIN) fiber of 25\,$\mu$m radius are shown in Fig.~\ref{fig:GRIN_step_compare} for a 1030\,nm wavelength and having the same index contrast. These fibers result in fundamental modes of a similar size, and are therefore straightforward to compare. For simplicity we say that the mode profiles are $M_i(x,y)$ where $i$ is a chosen index according to the convention that effective index decreases with mode number.

The first 6 modes are shown, which for these two fibers have exactly the same qualitative shape. The main difference is that in the GRIN fiber the higher-order modes begin to have larger wings (extending into the larger parabolic index profile), where the modes in the step-index fiber remain more tightly contained. The propagation parameter $\beta_0^{(i)}$ for each mode is related to the effective index, and determines how the modes propagate relative to each other. Higher-order modes in any waveguide will extend further into the cladding of the waveguide and therefore generally have a decreasing effective index.

The higher-order propagation parameters $\beta_n^{(i)}=\partial^n\beta_0^{(i)}/\partial\omega^n$ determine the group velocity ($n=1$) and the group velocity dispersion ($n=2$) and so on for each mode. When comparing the step-index and GRIN fiber, the step-index fiber has relatively continuously changing propagation parameters, where the GRIN fiber has groups of modes where the parameters are almost identical. Additionally, the GRIN fiber has a group velocity (and group velocity dispersion) that increases much more slowly with mode number, which provides a significant opportunity for interaction between the modes even after long propagation distances. The grouping and the magnitude of the group velocity difference can be seen on the right of Fig.~\ref{fig:GRIN_step_compare} where $\Delta\beta_1^{(i)}=\beta_1^{(i)}-\beta_1^{(1)}$ is shown for both fibers.

For simplicity, in the following sections we will mainly consider the case of the GRIN fiber shown in Fig.~\ref{fig:GRIN_step_compare}, but the qualitative behavior will be similar for any cylindrically symmetric multi-mode waveguide. Other waveguides are discussed in Section~\ref{sec:rect}.

\subsection{Basic coupling}
\label{sec:coupling}

A laser pulse-beam propagating in the $z$ direction with a Gaussian profile in space and time/frequency has an electric field $E=\tilde{A}\exp(-ikz)$ with a spectral amplitude~\cite{siegman86} 

\begin{align}
\begin{split}
\tilde{A}(x,y,z,\omega)=&\frac{\tilde{A}_0w_0}{w(z)}e^{-\left(\frac{x}{w(z)}\right)^2}e^{-\left(\frac{y}{w(z)}\right)^2}\\
&\times e^{i\left(\phi_G(z)-\frac{\omega\left(x^2+y^2\right)}{2cR(z)}\right)}e^{-\left(\frac{\delta\omega}{\Delta\omega}\right)^2},
\end{split}\\
w(z)=&w_0\sqrt{1+\left(z/z_R\right)^2}, \\
\phi_G=&\tan^{-1}\left({z}/{z_R}\right), \\
R(z)=&z+z_R^2/z,
\end{align}


\noindent where $w_0$ is the focused beam waist, with the focus at $z=0$, $\delta\omega=\omega-\omega_0$, $\Delta\omega$ is the frequency bandwidth ($\Delta\omega=2/\tau_0$, $\tau_0$ the Fourier-limited pulse duration), and $z_R=\omega_0w_0^2/2c$ is the Rayleigh range. The beam propagates in $z$, and $x$ and $y$ are the transverse spatial coordinates. Note that if we set $\omega=\omega_0$ in the curvature phase term, a suitable approximation when $z$ is small compared to $z_R$, unless the pulse is of few-cycle duration, then we can write the fields as separable functions $\tilde{A}(x,y,z,\omega)=p(x,y,z)\tilde{g}(\omega)$. Due to this separability the field will always have the same temporal term $g(t)=\exp{(i\omega_0{t}-(t/\tau_0)^2)}$, which is the Fourier-transform of $\tilde{g}(\omega)$. If $z$ is not small compared to $z_R$, then the beam develops a delay along with its curvature, such that the temporal function is $g(t-|r|^2/2cR(z))$. These nuances will be addressed at the end of this section and in the discussion section.

The coupling in the separable case to the mode $M_i$ of a fiber can be simply written as

\begin{align}
\begin{split}
&\eta_i=\\
&\frac{\left|\int\int \tilde{A}(x,y,\Delta{z},\omega)M_i(x,y)dxdyd\omega\right|^2}{\int\int|\tilde{A}(x,y,\Delta{z},\omega)|^2dxdyd\omega\int\int|M_i(x,y)|^2dxdy}
\end{split}\\
&=\frac{\left|\int\int p(x,y,\Delta{z})M_i(x,y)dxdy\right|^2}{\int\int|p(x,y,\Delta{z})|^2dxdy\int\int|M_i(x,y)|^2dxdy},
\end{align}

\noindent where the frequency dependence notably drops out (and it is commonplace to not be mentioned whatsoever)---the coupling is always purely a number such that the modes all have the same frequency content and therefore the same unmodified temporal profile. The offset between the waist position of the laser and the plane of the fiber facet is $\Delta{z}$. It has been known for decades that laser parameters can easily affect the overall coupling into multi-mode waveguides~\cite{niuJ07} and that this can modify linear and nonlinear processes~\cite{ahsan20}.

If the pulse were few-cycle, i.e. the bandwidth was comparable to the central frequency, then the field would no longer be separable and this equation would not strictly hold. Additionally, the mode profiles $M_i$ are also in reality frequency-dependent---this is part of how the higher-order propagation constants are calculated, as detailed in the previous section---but for the calculation of the coupling $\eta$ this dependence does not contribute significantly. If the pulse were few-cycle then this would also need to be taken into account. Lastly, it is important to note that $\eta_i$ is the coupling efficiency of \textit{energy} into each mode, that it is a real number, and the total coupling into the guided modes of the fiber is simply $\sum_i\eta_i$.

\begin{figure}[htb]
	\centering
	\includegraphics[width=\linewidth]{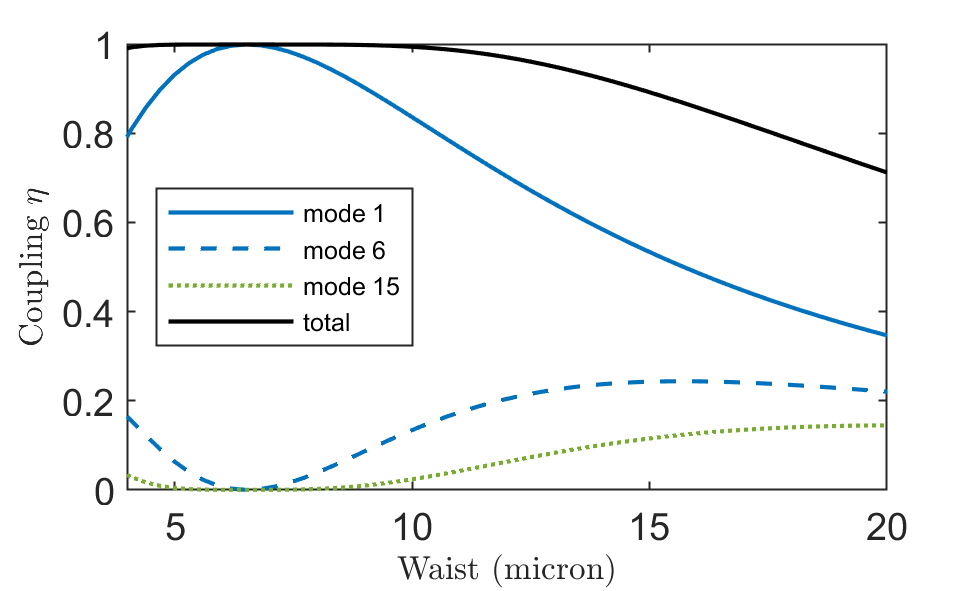}
	\caption{Coupling to the first three cylindrically symmetric modes of the canonical GRIN fiber as a function of the waist when alignment is perfect. The waist with the best coupling to the fundamental mode is roughly 7\,$\mu$m. }
	\label{fig:coupling_waist}
\end{figure}

A Gaussian beam incident on the GRIN fiber with perfect centering in $x$, $y$, and $z$ (i.e. $p(x,y,0)$) will have different coupling to only the cylindrically symmetric modes, mostly modes 1, 6, and 15 as shown in Fig.~\ref{fig:coupling_waist}. This is clear when understanding that a Gaussian itself is cylindrically symmetric, so the integral over all modes that are not cylindrically symmetric is zero. However, how much is coupled into each of modes 1, 6, and 15 depends on the waist $w_0$. Fig.~\ref{fig:coupling_waist} shows that there is a single waist where a Gaussian beam is best coupled to the fundamental mode of the GRIN fiber. In the case of the 25\,$\mu$m radius GRIN fiber considered, that ideal waist is roughly 7\,$\mu$m, and below and above that modes 6 and 15 are better coupled to. At larger waists the total coupling also eventually begins to decrease, due to the beam being significantly larger than the core.

\begin{figure}[htb]
	\centering
	\includegraphics[width=\linewidth]{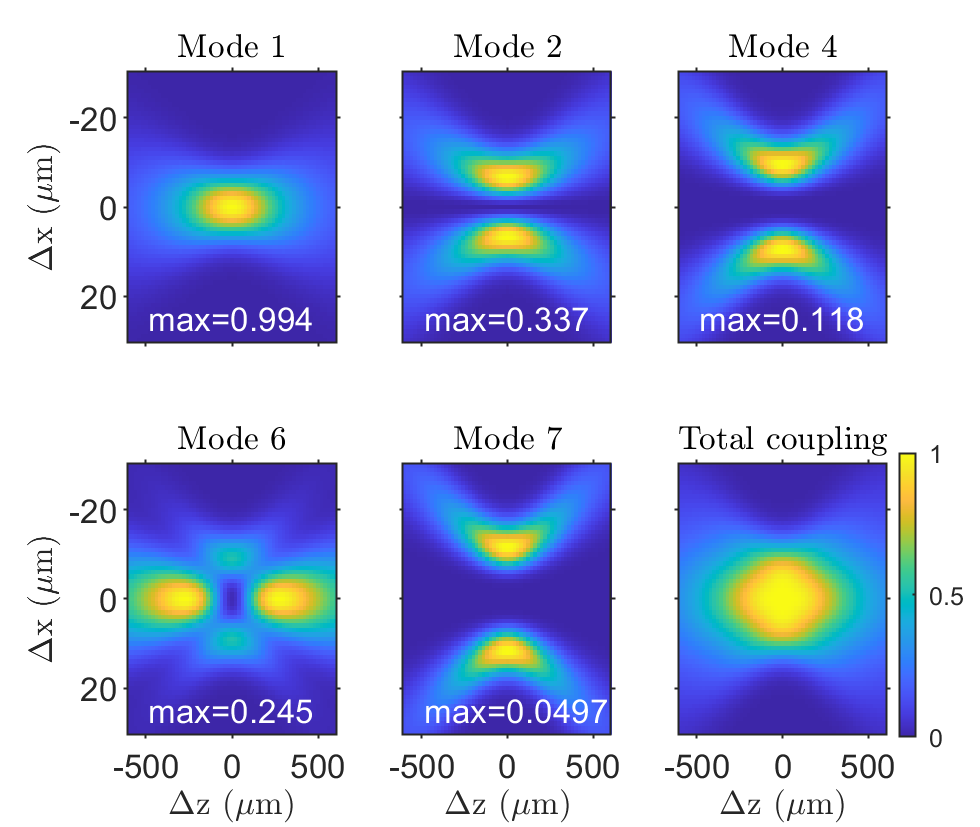}
	\caption{Coupling as a function of the $x$ and $z$ offset when the waist is ideal. Since there is no offset in $y$ there is zero coupling to modes that are odd in $y$ such as modes 3 and 5, so they are not shown. Optimal coupling is at $\Delta{x}=0$, $\Delta{z}\neq0$ for mode 6, and at $\Delta{z}=0$, $\Delta{x}\neq0$ for modes \{2,4,7\}.}
	\label{fig:coupling_offset}
\end{figure}

When a Gaussian beam is offset by $\Delta{x}$ with respect to the center of the fiber facet and $\Delta{z}$ from the facet itself, the field is proportional to $p(x-\Delta{x},y,\Delta{z})$. There will be significant coupling to various modes as shown in Fig.~\ref{fig:coupling_offset} with the ideal waist (7\,$\mu$m). Due to symmetry, modes that are odd in $y$ such as modes 3, 5, 8, etc. will have zero coupling regardless of the value of $\Delta{x}$ or $\Delta{z}$.

The coupling to the fundamental mode follows strongly the profile of the field amplitude itself, decreasing with an increase in either $|\Delta{x}|$ or $|\Delta{z}|$. The coupling to mode 6 is strongest when $\Delta{x}=0$ and $\Delta{z}\neq0$ such that the mode size is larger and there is nonzero spatial phase due to the curvature. For all of the other modes 2, 4, and 7, the behavior is similar: the peak coupling is when $\Delta{z}=0$ and $\Delta{x}\neq0$ and is symmetric with $\Delta{x}$ and $\Delta{z}$. The value of $\Delta{x}$ where the coupling is peaked increases with mode number for these three modes, and for all modes the peak energy coupled into the mode decreases with mode number (from mode 1 to 6, and from mode 2 to 4 to 7). Lastly, once either offset becomes large enough, the total energy coupled in starts to strongly decrease, since the beam is either too large or is missing the fiber core and energy does not couple to propagating modes.

The previous cases have considered variations in coupling to the different scalar modes of a GRIN fiber when changing either the waist or the offset (transverse or longitudinal) of the input focused Gaussian beam on the facet of the fiber. However, it is also possible to excite higher-order modes by adding a tilt of angle $\theta$, which is described by a linearly-varying spatial phase along one transverse coordinate $\exp{\left(i\theta{k}{x}\right)}$ ($k=\omega/c$). We make the same assumption made earlier with curvature that we can just consider the phase at the central frequency---$k\rightarrow{k_0}=\omega_0/c$---since the pulse is not few-cycle.

\begin{figure}[htb]
	\centering
	\includegraphics[width=\linewidth]{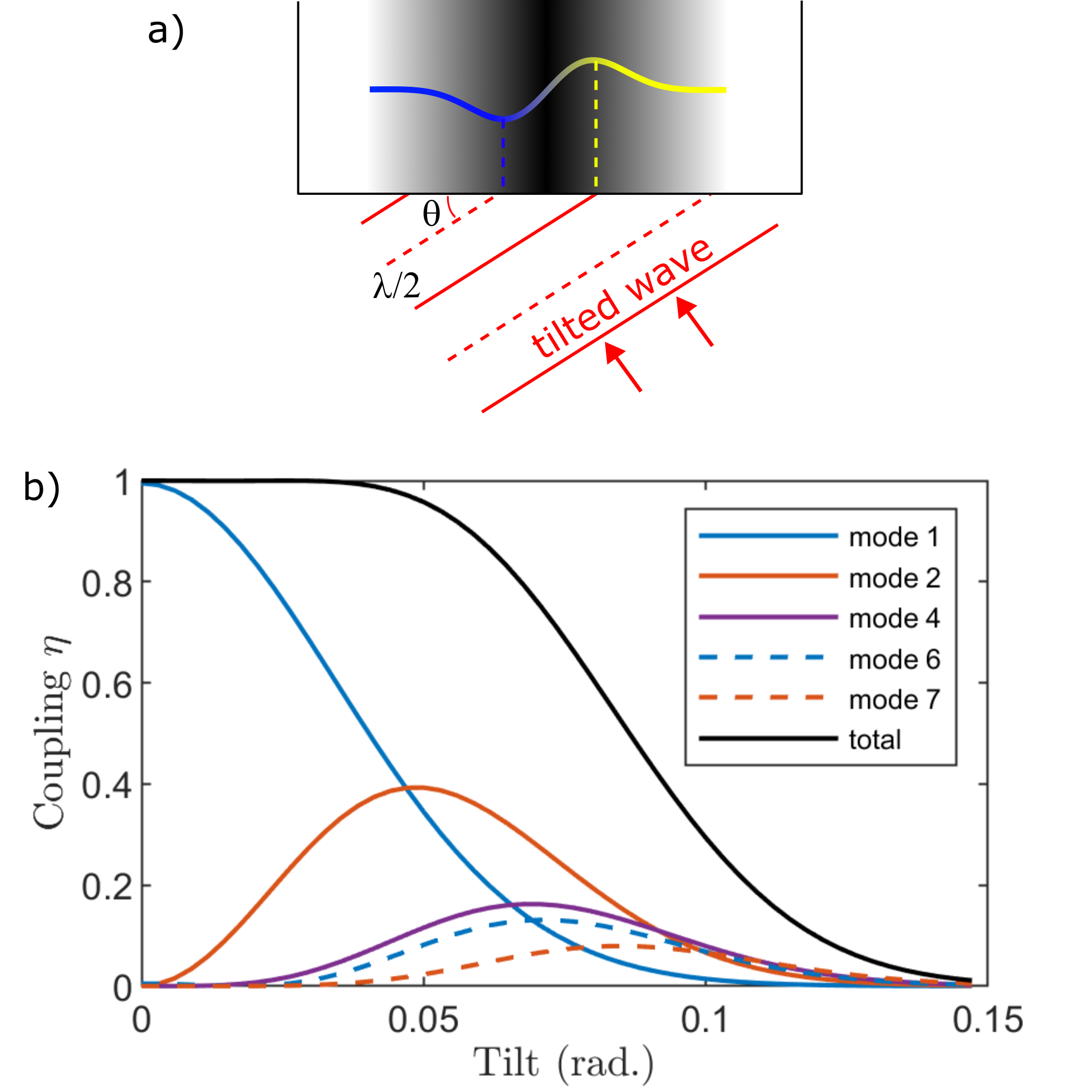}
	\caption{A tilted input pulse can couple preferentially to a higher-order mode (a) due to the phase difference across the input facet. Coupling as a function of tilt angle (b) to the canonical GRIN fiber when the alignment is perfect and the waist is ideal (7\,$\mu$m).}
	\label{fig:coupling_tilt}
\end{figure}

Looking specifically at mode 2 in Fig.~\ref{fig:GRIN_step_compare} and the coupling with tilt in Fig.~\ref{fig:coupling_tilt}(a), one can predict the coupling based on geometry. The profile of mode 2 reaches a maximum and a minimum at $\sim\pm5\,\mu$m (see Fig.~\ref{fig:GRIN_step_compare}). In order to couple well to that mode the phase caused by the tilt should produce a minimum and a maximum in the field at the same positions as for the mode, i.e. have a phase difference of $\pi$ between $x=\pm5\,\mu$m. This produces the relationship that $\theta=\arcsin{\left(\lambda_0/(4\times5\,\mu\textrm{m})\right)}$, a match for the peak coupling to mode 2 with 1030\,nm at$\sim0.05$\,radians shown in Fig.~\ref{fig:coupling_tilt}(b). The coupling can be better in mode 2 relative to mode 1 at larger angles, at the cost of the total energy coupled both into mode 2 and into all modes. The higher-order modes peak at a tilt that increases with the mode number, and with a peak coupling that decreases with mode number.

For both the curvature when there was a non-zero $\Delta{z}$ and in the case of tilt, we had to ignore a frequency-dependent term to keep the separability of the equations and to have coupling that did not depend on time or frequency. These were acceptable assumptions since we assumed that the pulses were not few-cycle. However, it is also important that the curvature or tilt not be too large. In the example of tilt this means that the time delay due to the tilt at an appropriate distance (for example the fiber radius $a$) is much less than the pulse duration: $a\tan{(\theta)}/c\tau_0\ll1$. For a 25\,$\mu$m radius fiber and a pulse of 200\,fs, this is a valid assumption until angles larger than 0.2\,radians, where energy does not couple to the fiber anymore. But for a 10 or 20\,fs pulse, this effect would matter. In the next section we will primarily discuss explicit spatio-temporal couplings, i.e. where an additional STC has been added to a pulse, on pulses of 200\,fs duration (emblematic of Yb-based laser sources that can operate at high-repetition rates). More nuanced cases for shorter pulses will be discussed more at the end of the manuscript.

As a final comment, beyond the basic control of offsets or beam sizes shown above, more arbitrary cases have been demonstrated. Spatial beam shaping has shown precise control of the coupling into very high-order modes of fibers~\cite{demas15-2}, and resulted in visible effects on nonlinear optical processes~\cite{demas15-1,demas17}.

\subsection{Modelling free-space beams with space-time couplings}
\label{sec:STCs}

When a free-space beam has a space-time coupling (STC), it is no longer possible to describe the pulse with separable functions dependent on only space and time respectively (nor space and frequency). The spectral amplitude was written $\tilde{A}(x,y,z,\omega)$ in the previous subsection, which could alternatively be written as $\tilde{A}(x,y,z,\omega)=p(x,y,z)\tilde{g}(\omega)$, where $\tilde{g}(\omega)=\exp{(-(\delta\omega/\Delta\omega)^2)}$. An STC is an aberration that causes a frequency-dependence in $p(x,y,z)$ or a spatial dependence in $\tilde{g}(\omega)$, and will naturally result in a similar unseparability when Fourier-transformed to time. There are some frameworks for analyzing STCs~\cite{kostenbauder90,akturk05}, but often the lack of separability results in the Fourier transform not being analytically solvable, such that only numerical integration can give the field in time.

As is discussed in many past works~\cite{bourassin-bouchet11,pariente16}, the significance of a given STC is increased for shorter and shorter pulses. Although a given STC will strictly exist on a pulse of any duration, its effect on propagation and evolution of the field will only matter for ultrashort pulses significantly shorter than 1 picosecond. A more nuanced treatment of this discussion is outside the scope of this work.

\section{Coupling to multi-mode waveguides with various space-time couplings}
\label{sec:STC_coupling}

Since an ultrashort pulse with STCs is no longer separable, the frequency dependence of the coupling will not necessarily drop out, so the coupling coefficient into a fiber will have a frequency dependence. We define a new parameter $\mu_i(\omega)$ which encapsulates this:

\begin{align}
\begin{split}
&\mu_i(\omega)=\\
&\frac{\int\int \tilde{A}(x,y,\Delta{z},\omega)M_i(x,y)dxdy}{\sqrt{\int\int|\tilde{A}(x,y,\Delta{z},\omega)|^2dxdyd\omega\int\int|M_i(x,y)|^2dxdy}}.
\end{split}
\end{align}

\noindent In contrast to $\eta_i$, $\mu_i(\omega)$ is the \textit{field} coupled into each mode (note the lack of the amplitude-squared operation in the numerator) and it is a complex number at each frequency, i.e. with an amplitude and a phase. Therefore, $\mu_i(\omega)$ is not an efficiency such as $\eta$, but rather the complex frequency envelope, and $\int|\mu_i(\omega)|^2d\omega$ gives the efficiency of the coupling into a certain mode. The phase has not been considered in previous work on in-coupling with STCs~\cite{guangZ18}. Not only can the phase vary with the mode number, but it can strongly vary with frequency. This means that there can be a significant (and different) spectral phase between the different modes depending on the field coupled in. Note that we still ignore the frequency-dependence of the modes $M_i$, since we consider only pulses that are composed of many cycles for the moment.

The temporal field coupled into each mode $f_i$ can be calculated by numerically Fourier-transforming $\mu_i$: $f_i(t)=\mathcal{F}_{\omega\rightarrow{t}}[\mu_i(\omega)]$. The spectral phase will strongly affect the temporal field and profile. We will consider various specific STCs and show how they affect the total coupling into the various modes of a GRIN fiber and how they affect the temporal profile coupled into those modes. The STCs introduced are all defined at the focus of the shaped ultrashort pulse-beam, i.e. where the coupling into the GRIN fiber takes place.

\subsection{Angular dispersion}
\label{sec:AD}

Angular dispersion is broadly defined to be when the tilt angle depends on frequency. This well known effect can be induced in free space by dispersive optical elements such as glass prisms or diffraction gratings. We consider here when an ultrashort pulse has angular dispersion at its focus. The complex spectral field amplitude $\tilde{A}_\textrm{AD}$ for a beam with angular dispersion (AD) along $x$ can be written generally as~\cite{martinez86}

\begin{align}
\begin{split}
\tilde{A}_\textrm{AD}(x,y,z,\omega)=&\frac{\tilde{A}_0w_0}{w(z)}e^{-\left(\frac{x}{w(z)}\right)^2}e^{-\left(\frac{y}{w(z)}\right)^2}e^{i\left(\frac{\tilde{\theta}\delta\omega{x}}{\Delta\omega{w_0}}\right)}\\
&\times e^{i\left(\phi_G(z)-\frac{\omega\left(x^2+y^2\right)}{2cR(z)}\right)}e^{-\left(\frac{\delta\omega}{\Delta\omega}\right)^2},
\end{split}
\end{align}


\noindent where all quantities are the same as before, and $\tilde{\theta}$ is a dimensionless quantity representing how strongly the tilt angle depends on frequency.

\begin{figure}[htb]
	\centering
	\includegraphics[width=\linewidth]{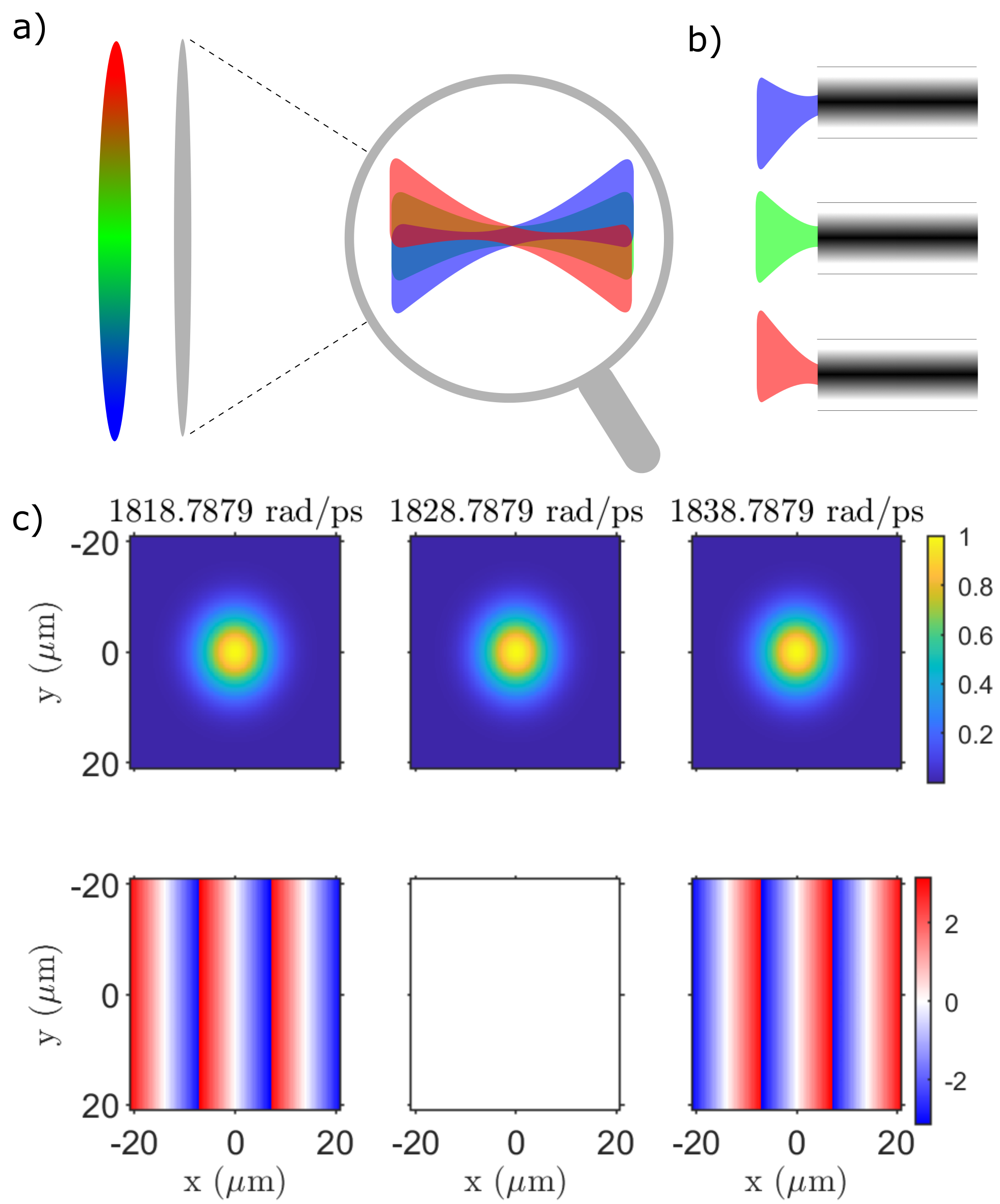}
	\caption{The concept of angular dispersion. An ultrashort pulse-beam that originally has spatial chirp is focused (a), where it develops angular dispersion. When coupled to a fiber facet (b) different colors will have a different input angle (spatial phase) while having essentially unaffected amplitude profiles (the angle is exaggerated for effect). This is quantitatively shown (c) for relevant frequencies of a 1030\,nm beam with a duration of 200\,fs ($w_0=7\,\mu$m, $\tilde{\theta}=\pi$) in the unvarying amplitude (top), but varying phase (bottom).}
	\label{fig:AD_concept}
\end{figure}

To connect this description to how it would be produced in experiments, a schematic is sketched in Fig.~\ref{fig:AD_concept}(a). To produce the AD in the focus the large collimated beam should actually have the colors separated in space (spatial chirp) along $x$. Then this spatial separation will be translated into angular separation in the focal plane of the lens. Rigorously, with this scheme, there should be a correction as well for the size of the beam along $x$, since frequencies besides $\omega_0$ are passing through the focal plane at an angle. But since we deal with very small angle we consider this correction to be small.

When the beam with AD impinges on the input facet of the GRIN fiber, sketched in Fig.~\ref{fig:AD_concept}(b), the amplitude will be centered for all frequencies, but the tilt angle will change significantly. This is shown quantitatively for the standard parameters of this section (1030\,nm wavelength, 200\,fs duration, 7\,$\mu$m waist) in Fig.~\ref{fig:AD_concept}(c) for three example frequencies.

\begin{figure}[htb]
	\centering
	\includegraphics[width=\linewidth]{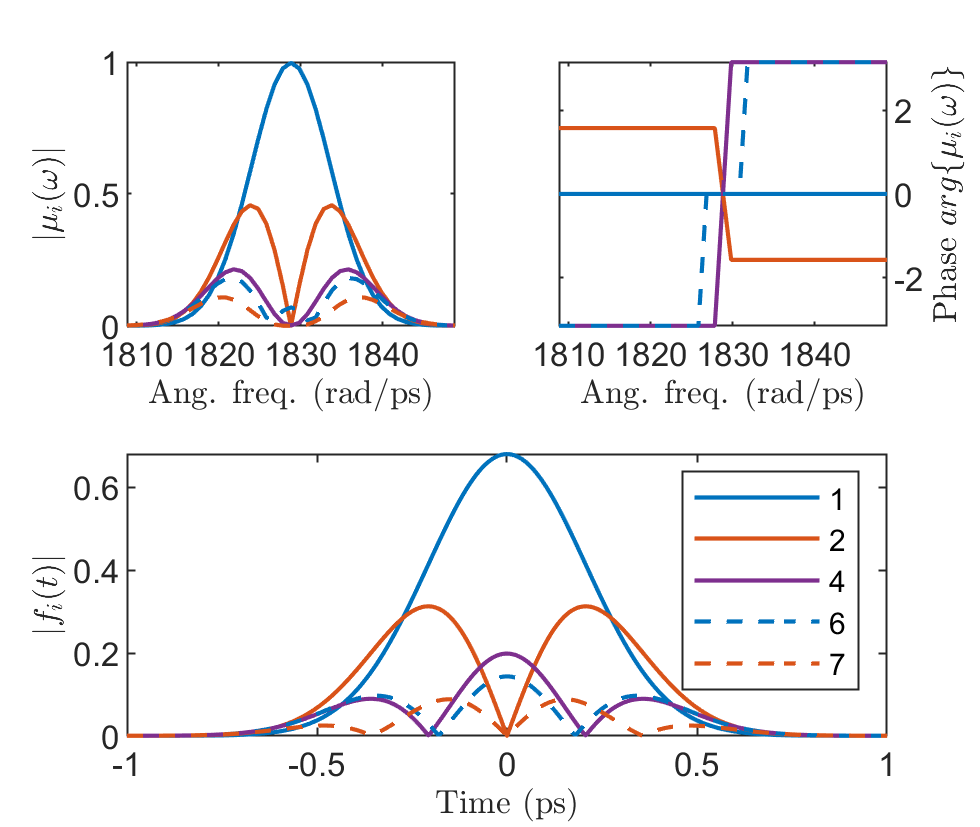}
	\caption{Coupling with angular dispersion for the canonical GRIN fiber when $w_0=7\,\mu$m, $\tau_0=200$\,fs, $\tilde{\theta}=\pi$, and $\Delta{z}=\Delta{x}=0$ (no offset).}
	\label{fig:AD_results}
\end{figure}

To predict the coupling into the GRIN fiber with AD, intuition can be used based on the coupling as a function of standard tilt described in Section~\ref{sec:coupling} and shown in Fig.~\ref{fig:coupling_tilt}. The frequencies that have a larger tilt, those further from $\omega_0$, will couple more to higher-order modes and eventually couple less overall. This is exactly what we see when calculating $\mu_i(\omega)$ and $f_i(t)$ for $\tilde{\theta}=\pi$ when there is no offset, shown in Fig.~\ref{fig:AD_results}.

At the central frequency $\omega_0$, there is only coupling to modes 1 and 6. At outer frequencies there is a larger coupling to modes 2, 4, and 7, with the coupling to mode 1 decreasing monotonically. The coupling to mode 6 drops to zero quickly, but peaks again both above and below $\omega_0$. The phase for mode 1 is zero at all frequencies, has a phase jump of $\pi$ for modes 2 and 7 when going from below $\omega_0$ to above it. The phase for mode 6 shows more complicated behavior, where it is zero near $\omega_0$, and jumps to $\pm\pi$ once it has passed it's zero coupling point. This combination of both amplitude and phase coupling as a function of frequency allows to calculate the temporal amplitude for each mode, shown on the bottom of Fig.~\ref{fig:AD_results}, which has not been shown in past works~\cite{guangZ18}.

\subsection{Spatial Chirp}
\label{sec:SC}

\begin{figure}[htb]
	\centering
	\includegraphics[width=\linewidth]{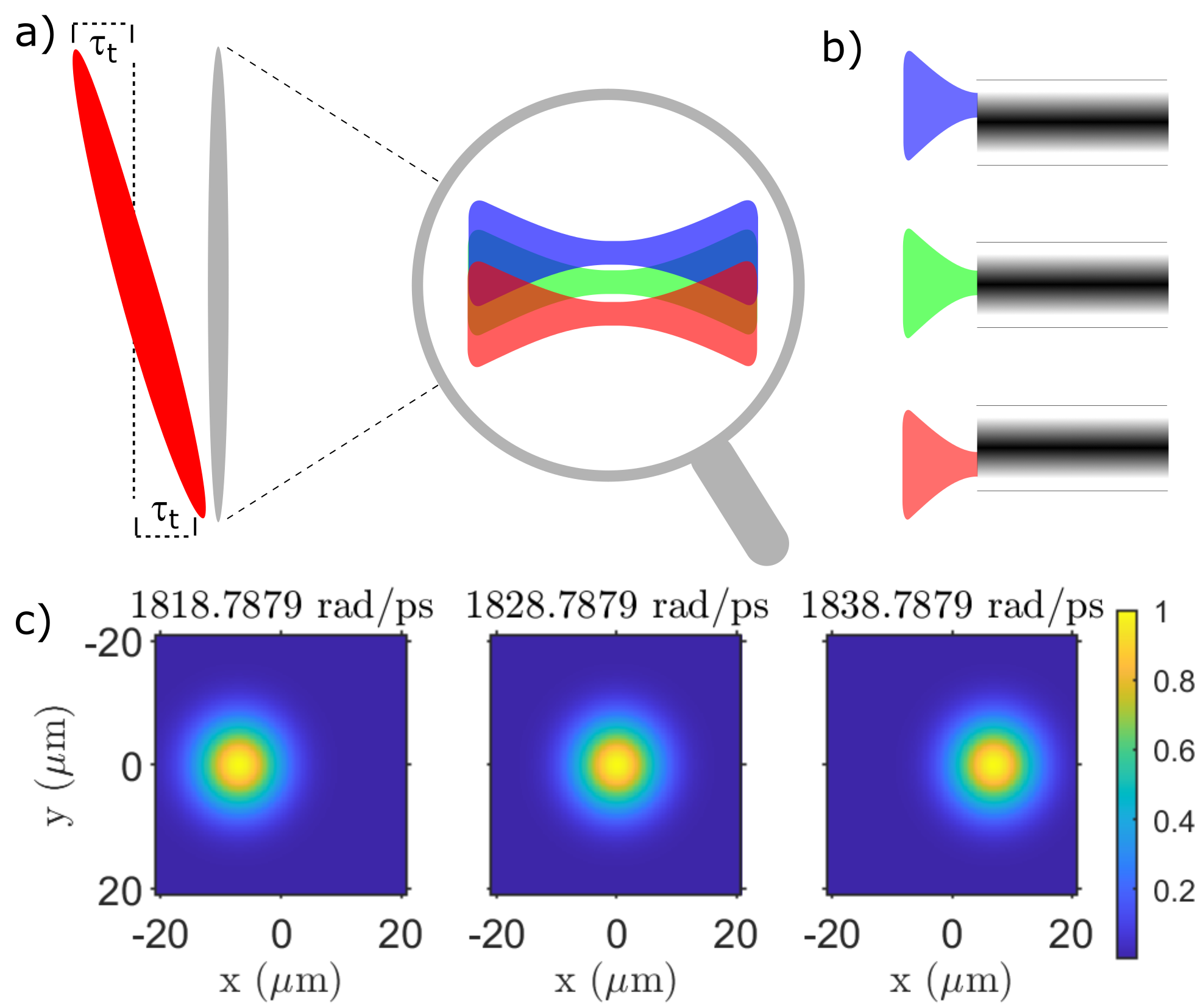}
	\caption{The concept of spatial chirp. An ultrashort pulse-beam that originally has pulse-front tilt (and therefore angular dispersion) is focused (a), where it develops spatial chirp. When coupled to a fiber facet (b) different colors with have a different offset while having essentially unaffected input angles (the offset is exaggerated for effect). This is quantitatively shown (c) for relevant frequencies of a 1030\,nm beam with a duration of 200\,fs ($w_0=7\,\mu$m, $\tau_t=\tau_0$) in the varying amplitude (constant and zero phase is not shown).}
	\label{fig:SC_concept}
\end{figure}

Spatial chirp (SC) is when the central transverse position of the different component frequencies depends on frequency~\cite{gu04}. This effect can notably be induced in free space by letting a beam with angular dispersion propagate (where there would be both AD and SC after propagation), or using a combination of prisms or gratings to subsequently remove the remaining AD. We consider here a beam with pure SC at its focus. It should be noted that this is equivalent to the temporal explanation of wavefront rotation that has applications to attosecond science~\cite{quere14,auguste16} and particle acceleration~\cite{thevenet19-2,mittelberger19}. The complex spectral field amplitude $\tilde{A}_\textrm{SC}$ for a beam with spatial chirp along $x$ can be written generally as

\begin{align}
\begin{split}
\tilde{A}_\textrm{SC}(x,y,z,\omega)=&\frac{\tilde{A}_0w_0}{w(z)}e^{-\left(\frac{x-x_0(\omega)}{w(z)}\right)^2}e^{-\left(\frac{y}{w(z)}\right)^2}\\
&\times e^{i\left(\phi_G(z)-\frac{\omega\left(\left[x-x_0(\omega)\right]^2+y^2\right)}{2cR(z)}\right)}\\
&\times e^{-\left(\frac{\delta\omega}{\Delta\omega}\right)^2},
\end{split}
\end{align}


\noindent with $x_0(\omega)=\tau_tw_0\delta\omega/2$ and $\phi_G(z)$, $w(z)$, and $R(z)$ as before. We choose to parameterize the SC using $\tau_t$, which is related to the magnitude of pulse-front tilt that would be needed on the collimated beam to produce the SC in the focused beam---sketched in Fig.~\ref{fig:SC_concept}(a). Pulse-front tilt on the collimated beam is equivalent to angular dispersion, which causes the different frequencies to focus to different transverse positions according to $x_0(\omega)$. This is precisely spatial chirp, and as sketched in Fig.~\ref{fig:SC_concept}(b) leads to some frequencies being better aligned transversely to the fiber core. More quantitatively, the field amplitude is shown in Fig.~\ref{fig:SC_concept}(c) for a beam with a waist of 7\,$\mu$m, a duration of 200\,fs, and $\tau_t=\tau_0$. Since the phase is flat at all frequencies it is not shown.

\begin{figure}[htb]
	\centering
	\includegraphics[width=\linewidth]{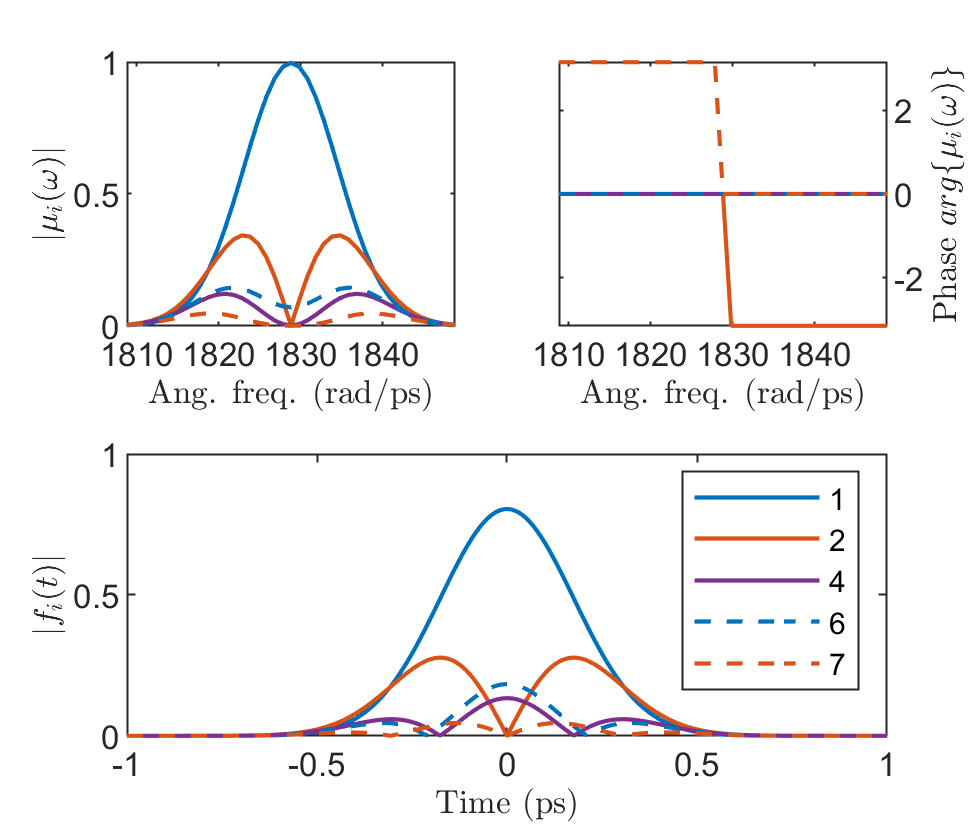}
	\caption{Coupling with spatial chirp for the canonical GRIN fiber when $w_0=7\,\mu$m, $\tau_0=200$\,fs, $\tau_t=\tau_0$, and $\Delta{z}=\Delta{x}=0$ (no offset).}
	\label{fig:SC_results}
\end{figure}

Intuition can be taken from the case of simple offset in Section~\ref{sec:coupling} and Fig.~\ref{fig:coupling_offset}, except that in the case of SC this will depend on frequency. The coupling for the same case as Fig.~\ref{fig:SC_concept}(c) is shown in Fig.~\ref{fig:SC_results} ($w_0=7\,\mu$m, $\tau_0=200$\,fs, $\tau_t=\tau_0$, and $\Delta{z}=\Delta{x}=0$) for the canonical 25\,$\mu$m radius GRIN fiber.

In fact, the coupling results with SC are very similar to those with AD, except that the overall coupling is lower. This makes sense since with SC the input field amplitude is less well aligned to the fiber core and therefore energy at the outer frequencies will not couple to the guided modes. Again the coupling at $\omega_0$ is purely to modes 1 and 6, and at outer frequencies becomes non-zero for modes 2, 4, and 7 (and higher-order modes that are even in $y$). The phase is such that for modes 2, 4, and 7 there is a $\pi$ difference between $\omega<\omega_0$ and $\omega>\omega_0$, and the phase is zero everywhere for mode 1 and mode 6 (contrary to the AD case).

\subsection{Longitudinal Chromatism}
\label{sec:LC}

\begin{figure}[htb]
	\centering
	\includegraphics[width=\linewidth]{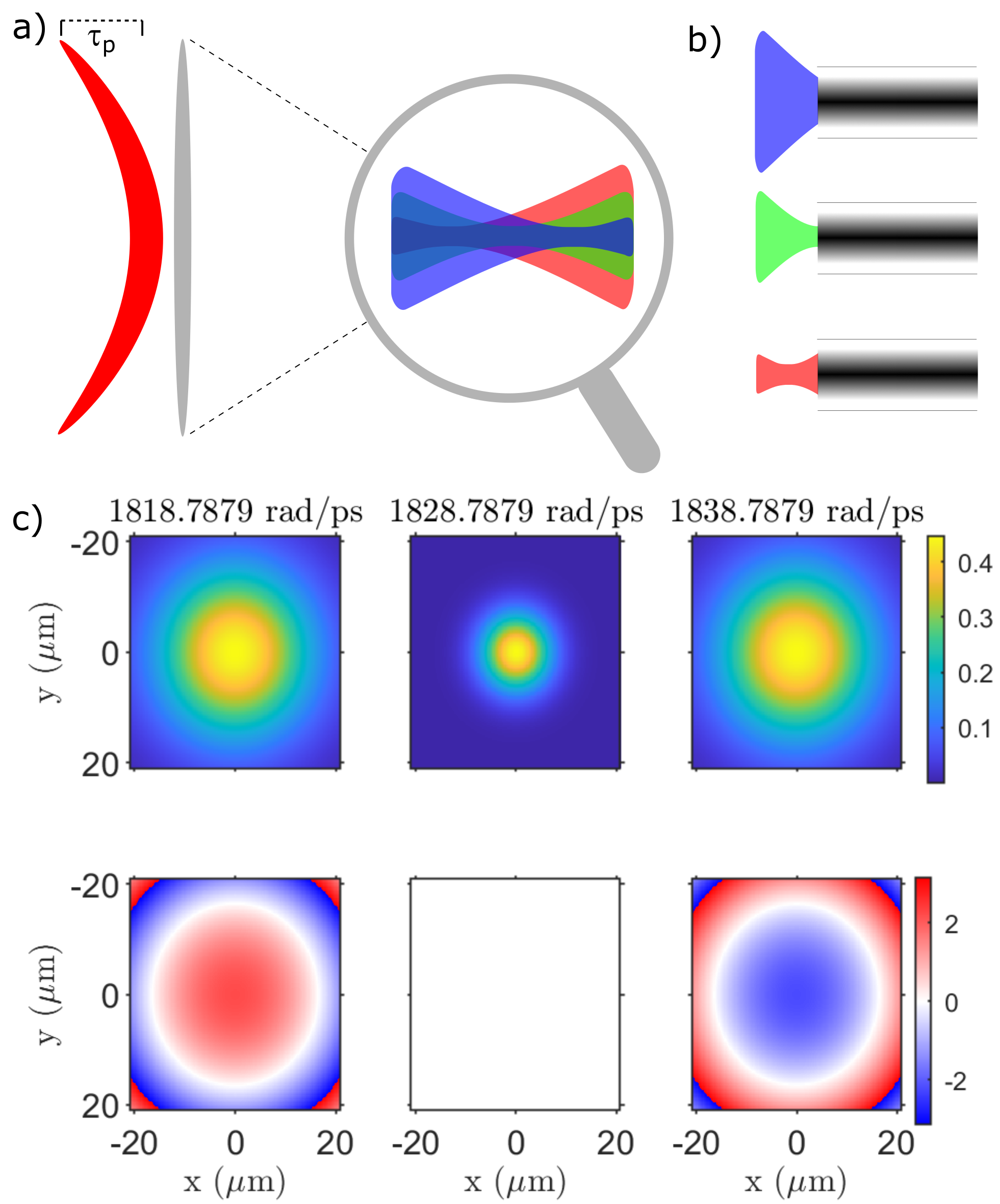}
	\caption{The concept of longitudinal chromatism. An ultrashort pulse-beam that originally has pulse-front curvature (and therefore chromatic curvature) is focused (a), where it develops longitudinal chromatism. When coupled to a fiber facet (b) different colors with have a different size and curvature (spatial phase) due to being focused at different $z$ positions (exaggerated for effect). This is quantitatively shown (c) for relevant frequencies of a 1030\,nm beam with a duration of 200\,fs ($w_0=7\,\mu$m, $\tau_p=\tau_0$) in the varying amplitude (top) and phase (bottom).}
	\label{fig:LC_concept}
\end{figure}

Longitudinal chromatism (LC) is another STC that describes when the longitudinal best-focus (or waist) position depends on frequency. This effect can be created by a highly chromatic singlet lens~\cite{bor88} or with a diffractive lens that is inherently chromatic~\cite{alonso18}, or with compound chromatic lens systems specifically designed for LC~\cite{sainte-marie17,jolly20-1}. The complex spectral field amplitude $\tilde{A}_\textrm{LC}$ for a beam with longitudinal chromatism along can be written generally as

\begin{align}
\begin{split}
\tilde{A}_\textrm{LC}(x,y,z,\omega)=&\frac{\tilde{A}_0w_0}{w(z,\omega)}e^{-\left(\frac{x}{w(z,\omega)}\right)^2}e^{-\left(\frac{y}{w(z,\omega)}\right)^2}\\
&\times e^{i\left(\phi_G(z,\omega)-\frac{\omega\left(x^2+y^2\right)}{2cR(z,\omega)}\right)}\\
&\times e^{-\left(\frac{\delta\omega}{\Delta\omega}\right)^2},
\end{split}\\
w(z,\omega)=&w_0\sqrt{1+\left(\left[z-z_0(\omega)\right]/z_R^2\right)^2},\\
\phi_G(z,\omega)=&\tan^{-1}\left(\left[z-z_0(\omega)\right]/{z_R}\right),\\
R(z,\omega)=&\left[z-z_0(\omega)\right]+z_R^2/\left[z-z_0(\omega)\right],
\end{align}


\noindent where we again choose to parameterize the effect with a characteristic time $\tau_p$, with $z_0(\omega)=\tau_p{z_R}\delta\omega$. This $\tau_p$ related to the magnitude of pulse-front curvature on the beam before focusing---sketched in Fig.~\ref{fig:LC_concept}(a). Since it is now the longitudinal best-focus position $z_0$ that depends on frequency, the waist, Guoy phase, and curvature all now have their own frequency dependence.

In the focus the different frequencies will have different beam sizes according to $w(z,\omega)$ and phase curvatures according to $1/R(z,\omega)$. A sketch of this is shown in Fig.~\ref{fig:LC_concept}(b) with $z=0$, where the central frequency is at its waist on the fiber facet, and the other frequencies above and below $\omega_0$ are both equally larger, but with opposite curvatures. This same case is shown quantitatively in Fig.~\ref{fig:LC_concept}(c) for a beam with a waist of 7\,$\mu$m, a duration of 200\,fs, and $\tau_p=\tau_0$.

\begin{figure}[htb]
	\centering
	\includegraphics[width=\linewidth]{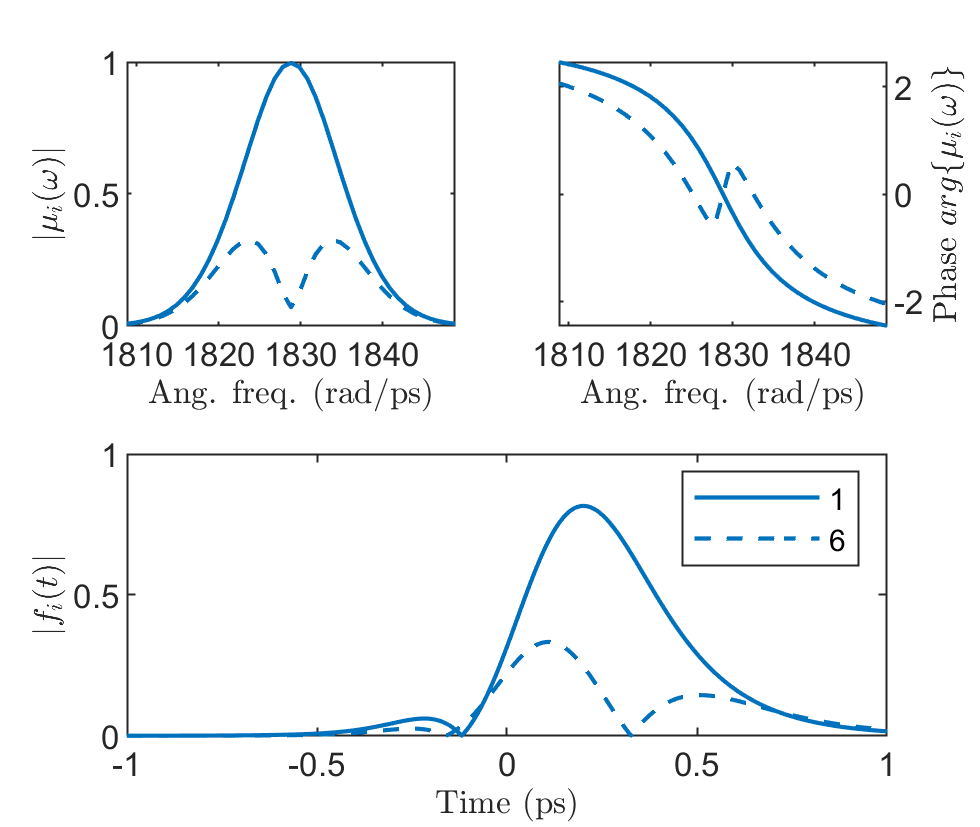}
	\caption{Coupling with longitudinal chromatism for the canonical GRIN fiber when $w_0=7\,\mu$m, $\tau_0=200$\,fs, $\tau_p=\tau_0$, and $\Delta{z}=\Delta{x}=0$ (no offset).}
	\label{fig:LC_results}
\end{figure}

The important consideration for the coupling is that, with LC, the beamsize and curvature will depend on frequency, but in a constrained way (i.e. not freely). Since the Gouy phase also depends on frequency, there will be an additional effect on the spectral phase in the different modes (but not on the coupled amplitude). And since the beam is cylindrically symmetric at all frequencies, the coupling with zero transverse offset will be only to cylindrically symmetric modes (modes 1 and 6). The coupling for the same case as Fig.~\ref{fig:LC_concept}(c) is shown in Fig.~\ref{fig:LC_results} ($w_0=7\,\mu$m, $\tau_0=200$\,fs, $\tau_t=\tau_0$, and $\Delta{z}=\Delta{x}=0$) for the canonical 25\,$\mu$m radius GRIN fiber.

The amplitude coupling with LC is not necessarily surprising, where the coupling to mode 1 decreases monotonically away from $\omega_0$, and peaks and then decreases for mode 6. However, the phase is especially different in the case of LC. Both mode 1 and mode 6 have a spectral phase that looks similar to $\tan^{-1}{(\delta\omega)}$, but for mode 6 there is a discontinuous section near $\omega_0$. This shape causes both mode 1 and mode 6 to loose time reversal symmetry (due to the cubic component of the spectral phase), and for mode 6 to have additional complexity. These results with LC provide a nice contrast to the cases of AD and SC, which were rather similar.

\subsection{Chromatic Astigmatism}
\label{sec:CAstig}

The last STC that we introduce is deemed chromatic astigmatism (CA), and has only been recently described theoretically for the first time~\cite{jolly23-1}. CA is a natural extension of LC, where the cylindrical symmetry is no longer present, but there is rather a saddle-like symmetry. More simply put, the longitudinal best-focus shift $z_0$, now dependent on a new characteristic time $\tau_a$, is chosen to have the opposite effect in $y$ than in $x$. Although CA has never been demonstrated experimentally, we believe it could be produced from optical systems with chromatic cylindrical lenses, and we are in the process of demonstrating this.

The complex spectral field amplitude $\tilde{A}_\textrm{CA}$ for a beam with chromatic astigmatism can be written as

\begin{align}
\begin{split}
\tilde{A}_\textrm{CA}(x,y,z,\omega)=&\frac{\tilde{A}_0w_0}{\sqrt{w_-(z,\omega)w_+(z,\omega)}} \\
&\times e^{-\left(\frac{x}{w_-(z,\omega)}\right)^2}e^{-\left(\frac{y}{w_+(z,\omega)}\right)^2} \\
&\times e^{i\left(\phi_{G,-}(z,\omega)-\frac{\omega x^2}{2cR_-(z,\omega)}\right)} \\
&\times e^{i\left(\phi_{G,+}(z,\omega)-\frac{\omega y^2}{2cR_+(z,\omega)}\right)}\\
&\times e^{-\left(\frac{\delta\omega}{\Delta\omega}\right)^2},
\end{split}
\end{align}


\noindent with $z_0(\omega)=\tau_a{z_R}\delta\omega$ and

\begin{align}
w_{\pm}(z,\omega)=&w_0\sqrt{1+\left(\left[z{\pm}z_0(\omega)\right]/z_R\right)^2},\\
\phi_{G,\pm}(z,\omega)=&\frac{1}{2}\tan^{-1}\left(\left[z{\pm}z_0(\omega)\right]/{z_R}\right),\\
R_{\pm}(z,\omega)=&\left[z{\pm}z_0(\omega)\right]+z_R^2/\left[z{\pm}z_0(\omega)\right].
\end{align}

\noindent Note the similarities to the LC case, except there are now separate waist, Gouy phase, and curvature terms for $x$ and $y$ that contain $z_0$ with an opposite sign.

\begin{figure}[htb]
	\centering
	\includegraphics[width=\linewidth]{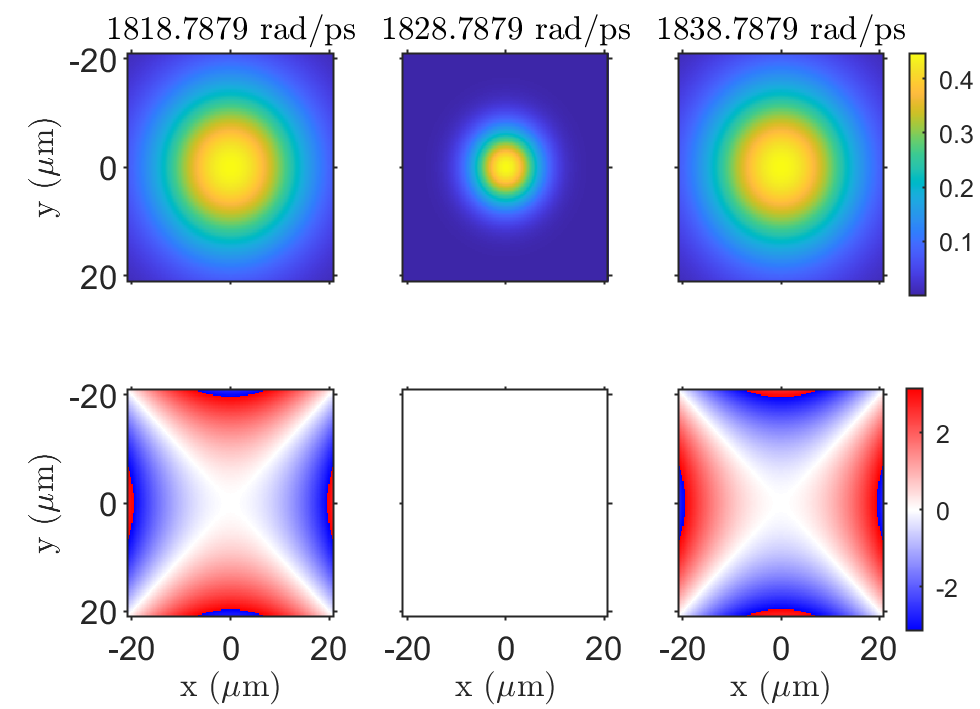}
	\caption{The concept of chromatic astigmatism is essentially the same as for longitudinal chromatism, except that the behavior in $x$ and $y$ is opposite---cylindrical symmetry is lost. Both the beam size (top) and spatial phase (bottom) is shown for relevant frequencies of a 1030\,nm beam with a duration of 200\,fs ($w_0=7\,\mu$m, $\tau_a=\tau_0$). Note that along $x$ the behavior is the same as for longitudinal chromatism (up to a constant in phase).}
	\label{fig:CA_input}
\end{figure}

The characteristics of an ultrashort pulse with CA are described in more detail in Ref.~\cite{jolly23-1}, but we can already see an example of the behavior with $z=0$ in Fig.~\ref{fig:CA_input}. The amplitude for all three frequencies is the same as with LC---all three frequencies have a round amplitude profile, and at $\omega_0$ the beam is smaller while at the two outside frequencies the beam is larger. However, the phase is different with the case of CA. Along the x-axis the phase is the same as with LC (up to a constant), but along $y$ the phase is inverted. Because of the saddle symmetry the phase is constant at $x=y=0$ and in general when $|x|=|y|$.

\begin{figure}[htb]
	\centering
	\includegraphics[width=\linewidth]{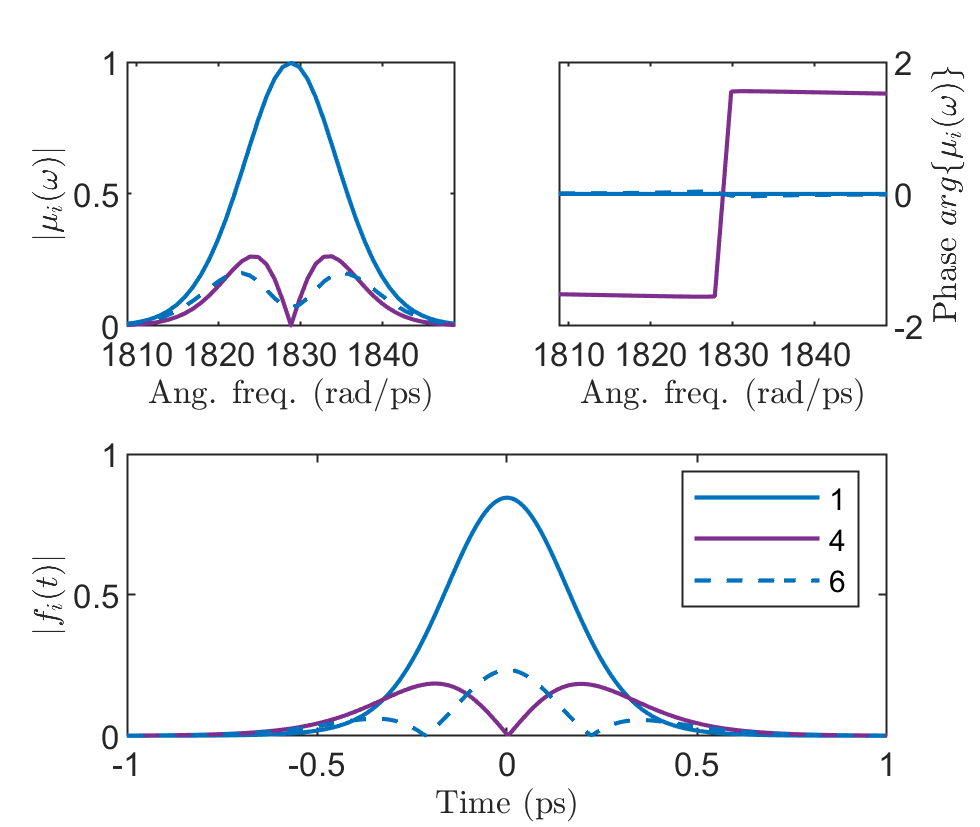}
	\caption{Coupling with chromatic astigmatism for the canonical GRIN fiber when $w_0=7\,\mu$m, $\tau_0=200$\,fs, $\tau_a=\tau_0$, and $\Delta{z}=\Delta{x}=0$ (no offset).}
	\label{fig:CA_results}
\end{figure}

The coupling for the same case as Fig.~\ref{fig:CA_input} is shown in Fig.~\ref{fig:CA_results} ($w_0=7\,\mu$m, $\tau_0=200$\,fs, $\tau_a=\tau_0$, and $\Delta{z}=\Delta{x}=0$) for the canonical 25\,$\mu$m radius GRIN fiber. Interestingly, there is only coupling to modes 1, 4, and 6. This is the first coupling that has had coupling to mode 4 while not having coupling to modes 2 or 3 (i.e. the second-lowest mode that is coupled to is mode 4). The coupling to mode 1 decreases monotonically away from $\omega_0$, and for both mode 4 and mode 6 it peaks and then decreases away from $\omega_0$. In contrast to LC and similar to AD and SC, the phase is once again discrete jumps. In fact for modes 1 and 6 the phase is zero, but for mode 4 it is a jump of $\pi$ from below $\omega_0$ to above it. Due to this difference in phase, despite having similar amplitude coupling, modes 4 and 6 have different coupling in the time domain (mode 6 is more complex than mode 4).

\begin{figure}[htb]
	\centering
	\includegraphics[width=\linewidth]{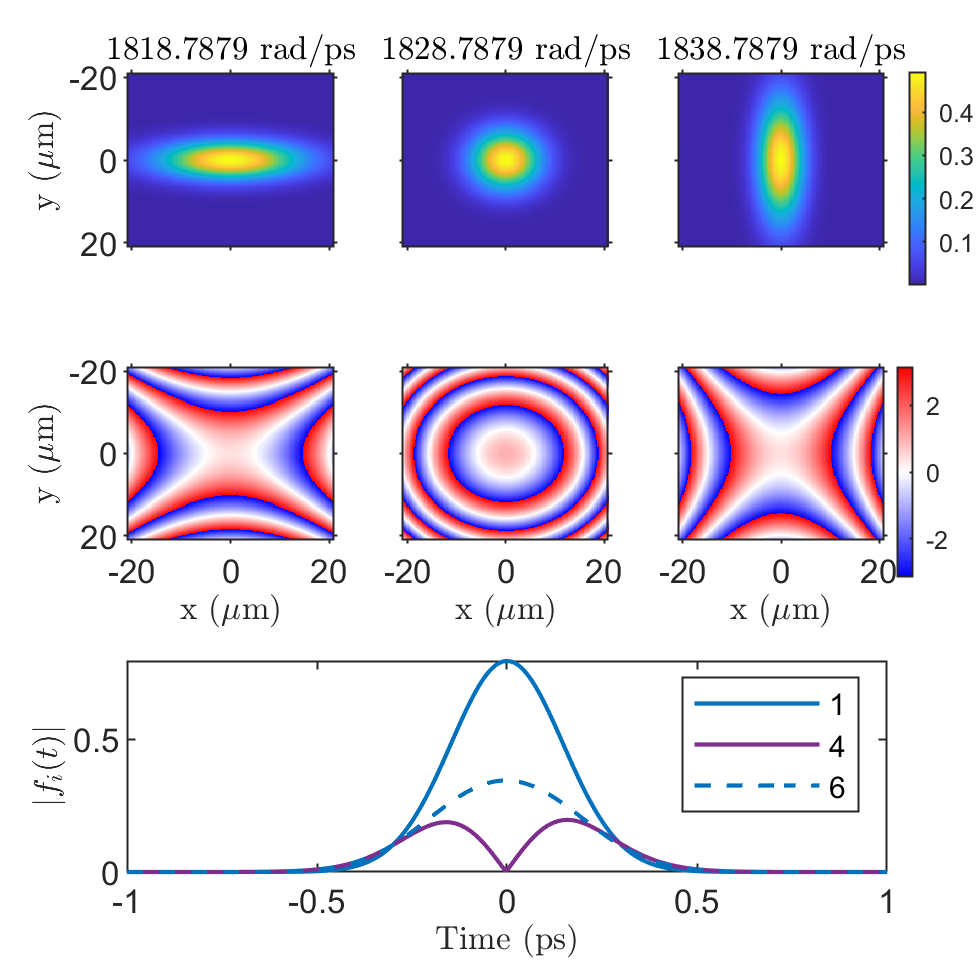}
	\caption{Coupling with chromatic astigmatism for the canonical GRIN fiber when $w_0=4\,\mu$m, $\tau_0=200$\,fs, $\tau_a=\tau_0$, $\Delta{z}=70\,\mu$m, and $\Delta{x}=0$. With the smaller waist at the focus, $\Delta{z}$ is such that at $\omega_0$ the beam waist is 7\,$\mu$m as before but now with curvature, and there is significant asymmetry in the amplitude and phase away from $\omega_0$. There is then a stronger coupling to both modes 4 and 6, but with a different symmetry.}
	\label{fig:CA_results2}
\end{figure}

Chromatic astigmatism for a different case with offset in $z$ ($w_0=4\,\mu$m, $\tau_0=200$\,fs, $\tau_a=\tau_0$, $\Delta{z}=70\,\mu$m, and $\Delta{x}=0$) is shown in Fig.~\ref{fig:CA_results2}. In this case the input beam field is significantly different for the different frequencies. The amplitude is still round at $\omega_0$, but the phase is as for a diverging beam. Above and below $\omega_0$ the beam amplitude becomes elliptical with the phase being an asymmetric saddle, where above and below are equal to each other when inverting $x$ and $y$.

The coupling in this second case of CA also only excites mode 1, 4, and 6. It can be seen in the symmetry of the modes and $A_\textrm{CA}$ that the only modes with non-zero coupling will be modes that are even in both $x$ and $y$ (when there is zero transverse offset). However, in this second case the coupling to mode 6 is peaked at $t=0$ and it decreases monotonically, such that mode 4 is more complex than mode 6 in the time domain.

\section{Discussion}
\label{sec:discussion}

We have now built up the motivation and basic tools for space-time coupling into multi-mode fibers, and shown the frequency and temporal amplitudes coupled into the individual modes (up to mode 7 to be compact) for four different STCs that are either commonplace or relatively simple to visualize and create experimentally. This has already been a big step, but of course there are important nuances that can still be tackled and other important scenarios that we have so far not sketched. For example, all of the STCs in the previous section were linear in frequency (i.e. containing only the term linear in $\delta\omega$). Theoretically each of the cases could have terms of arbitrary order in frequency. We discussed that the couplings discussed result in space-dependent spectral phase, but spatially-homogeneous spectral phase could also be added to any input pulse (second-order GDD, third-order TOD, etc.). In the following we will address a few other nuances that may be interesting.

\subsection{Combining various STCs}
\label{sec:combination}

A simple extension of the four individual STCs touched upon in Section~\ref{sec:STC_coupling} is the coupling to the modes of a multi-mode waveguide with a combination of two or more of the same couplings. For simplicity we will consider only spatial chirp, longitudinal chromatism, and chromatic astigmatism. SC and AD produce relatively similar results, and to produce AD in the focus requires a spatially chirped beam before focusing (Fig.~\ref{fig:AD_concept}(a)), which none of the other couplings do. The individual parameters $\tau_t$, $\tau_p$, and $\tau_a$ for SC, LC, and AC respectively, act on different parts of the field equation, so at least theoretically can simply be combined.

\begin{figure}[htb]
	\centering
	\includegraphics[width=\linewidth]{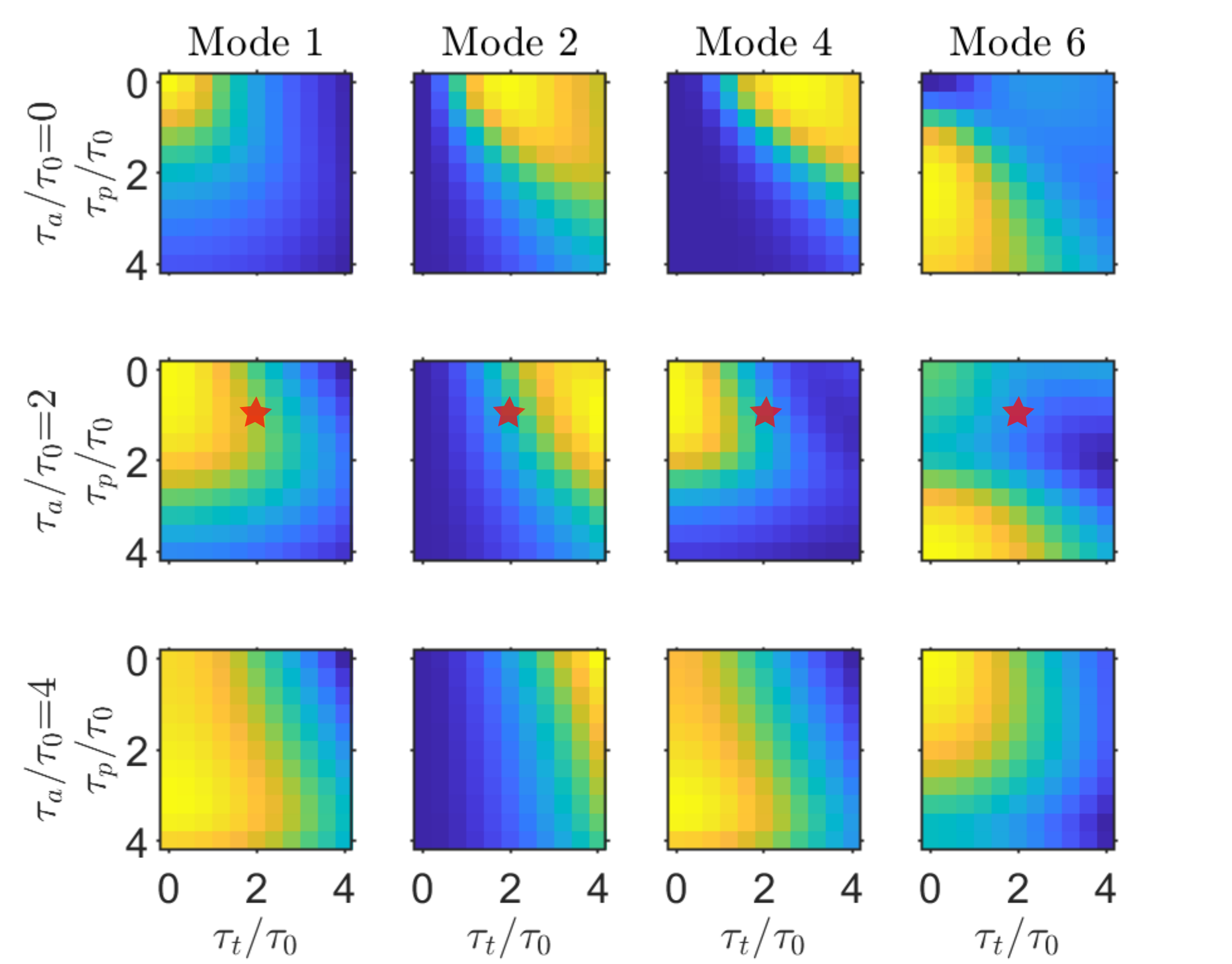}
	\caption{Total coupling ($\int|\mu_i|^2 d\omega$) to four example modes with a combination of SC, LC, and CA ($\tau_t$, $\tau_p$, and $\tau_a$, respectively) for the canonical GRIN fiber. Other parameters are $w_0=7\,\mu$m, $\tau_0=200$\,fs, and $\Delta{z}=\Delta{x}=0$ (no offset). The red stars denote the results shown in Fig.~\ref{fig:combination2}.}
	\label{fig:combination}
\end{figure}

In Fig.~\ref{fig:combination} the total energy coupled to each of modes 1, 2, 4, and 6 is shown for a range of $\tau_t$, $\tau_p$, and $\tau_a$ values. Characteristics shown before can be seen, where for example mode 2 can only be coupled to with SC, or mode 6 can only be coupled to well with $\tau_p$ or $\tau_a$. With combinations of the different couplings, however, there can be significant energy in all of the modes. We considered only the total energy to more easily visualize the results, but of course each point has a different temporal phase and profile.

\begin{figure}[htb]
	\centering
	\includegraphics[width=\linewidth]{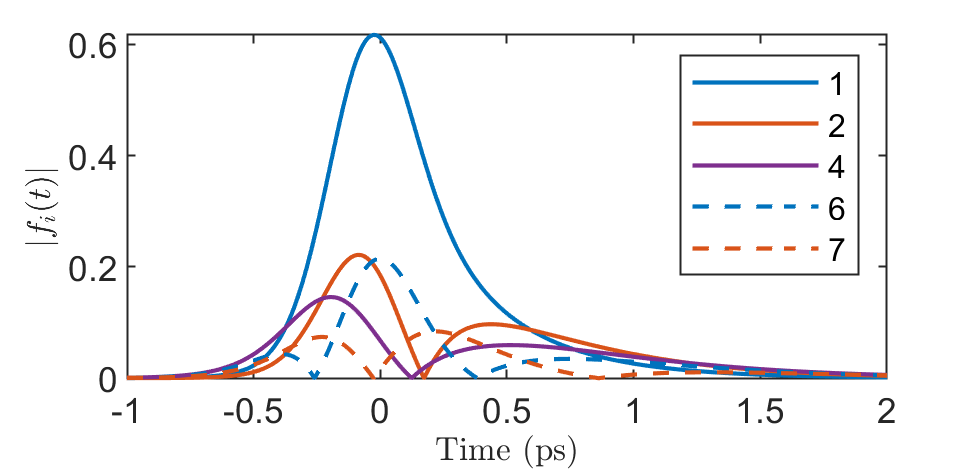}
	\caption{Example of the temporal envelope coupled to the canonical GRIN fiber when $\tau_t=2\tau_0$, $\tau_p=\tau_0$, and $\tau_a=2\tau_0$ (the red stars in Fig.~\ref{fig:combination}).}
	\label{fig:combination2}
\end{figure}

Fig.~\ref{fig:combination2} shows the temporal envelope coupled to various modes at one of the points from Fig.~\ref{fig:combination}, when $\tau_t=2\tau_0$, $\tau_p=\tau_0$, and $\tau_a=2\tau_0$. There is a similar overall trend to the case of pure SC, except that there is now a lack of temporal symmetry and an overall higher amplitude in the higher-order modes. Using a combination of the simple STCs allows to more finely tune the input temporal field to a desired mode or set of modes and their temporal profiles.

\subsection{Arbitrary space-time input}
\label{sec:arbitrary}

The previous section, using a combination of simple STCs, immediately raises the question of arbitrary input to a multi-mode waveguide. If we have a desirable initial distribution in the waveguide, i.e. a known complex amplitude for each mode $\mu_i(\omega)$ that is best for a given application, found using some optimization algorithm or inverse design, how can that be achieved?

The reason that simpler low-order STCs were used up to this point is that they are easier to model with a single parameter, but also that they are simpler to implement experimentally---generally with a single dispersive optic of appropriate material and dimensions. However, of course this is not the only method to do space-time shaping. In recent years there have been multiple breakthroughs in space-time shaping to achieve almost arbitrary control, while improving the fidelity of such control~\cite{mounaix20,yessenov22-1,caoQ22,cruz-delgado22,chenL22}. With such arbitrary space-time shaping the targeted complex amplitude in the multi-mode waveguide could be made, where the additional step is essentially the calculations done in this work that find the coupling to the different modes from a free-space input.

This topic is relatively open-ended, and because there are not yet mature applications for tailored multi-mode complex fields we won't address it in-depth. However, such consideration will surely be part of a future design procedure.

\subsection{Towards few-cycle pulses}
\label{sec:few-cycle}

As has already been alluded to in previous sections, there are numerous assumptions that no longer hold when considering few-cycle or single-cycle ultrashort pulses. Indeed applications often consider nonlinear optics in fibers to create shorter pulses~\cite{carlson19}, and the input pulses are therefore much longer than of few-cycle duration. Still, we believe that addressing these nuanced issues is important for such a tutorial article.

A pulse is few-cycle when the Fourier-limited duration is only a few times longer than the optical cycle, i.e. $\tau_0\sim2\pi/\omega_0$. An equivalent condition is that the bandwidth is comparable to the central frequency, i.e. $\Delta\omega\sim\omega_0$. If one ignores small factors then these conditions are the same (knowing $\Delta\omega=2/\tau_0$).

There are a few steps necessary to properly model ultrashort pulses that are of few-cycle duration or shorter, without yet considering coupling into a waveguide. First of all, in general the $\omega$ factor in the curvature phase term of Eq.~1 cannot be replaced with $\omega_0$, except for at $z=0$ where the term is zero in any case. This is because the time delay due to curvature may be comparable to the pulse duration, even at small $z$. Additionally, the frequency-dependence of $z_R$ or $w_0$ must be considered. The true relationship is $z_R=\omega w_0^2/2c$ or $w_0=\sqrt{2 c z_R/\omega}$, so even if one of $z_R$ or $w_0$ is frequency-independent, the other will have frequency-dependence that matters for a few-cycle pulse~\cite{porras99}. In fact, depending on the generation method, the specific nature of the frequency dependence can be parametrized by a factor $g_0$~\cite{porras09} which represents the linear slope of the frequency dependence of the Rayleigh range: $g_0=-(dz_R/d\omega)_{@\omega_0}(\omega_0/z_{R0})$. This has been measured for representative few-cycle systems~\cite{hoff17-1} and shown to have a significant effect on some femtosecond time-scale processes~\cite{hoff17-2,zhangY20,jolly20-2}.

Lastly, to avoid erroneously including zero or negative frequencies, the spectral envelope can be more rigorously modeled by, for example, a Poisson-like distribution instead of a Gaussian~\cite{caron99}. This type of spectral envelope goes strictly to zero as $\omega$ goes to zero regardless of the spectral bandwidth.

Once the above are considered in the modeling of an input ultrashort pulse focusing in a vacuum, the last ingredient is related to the waveguide itself. The dependence of the modal size (and shape) on frequency was not considered in our calculations of the couplings without STCs ($\eta_i$) or with STCs ($\mu_i(\omega)$). To be strictly correct in the case of a few-cycle pulse, this should be considered.

\begin{figure}[htb]
	\centering
	\includegraphics[width=\linewidth]{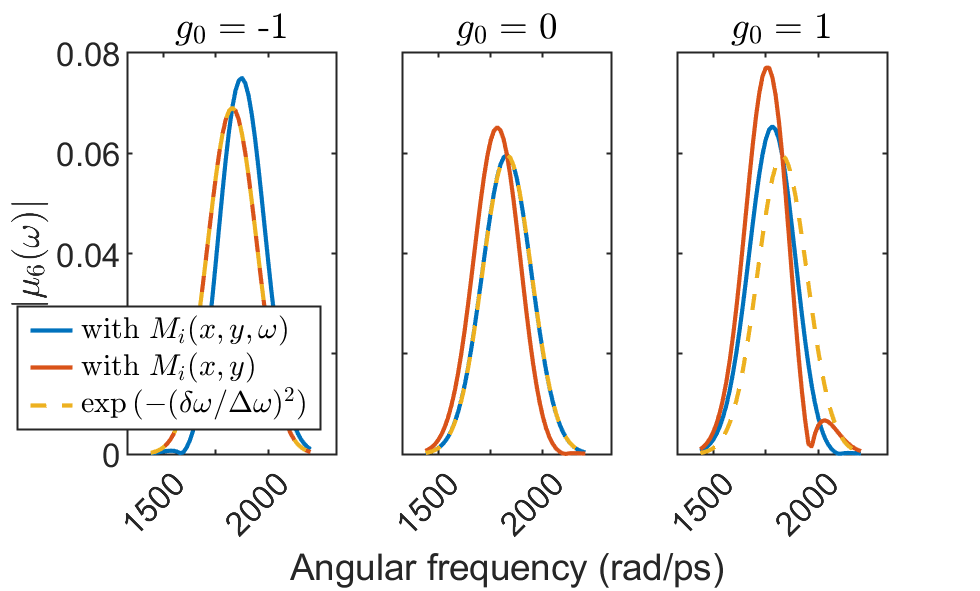}
	\caption{Example of effects in a few-cycle pulse. 12\,fs pulses with $g_0=-1$, 0, and +1 are coupled to the canonical 25\,$\mu$m GRIN fiber, and the coupling to mode 6 is shown with $\Delta{x}=\Delta{z}=0$. The coupling is different when considering frequency-dependence of the mode profiles or not.}
	\label{fig:few-cycle}
\end{figure}

In Fig.~\ref{fig:few-cycle} the coupling of 12\,fs pulses with $w_0(\omega_0)=7\,\mu$m to mode 6 of the same GRIN fiber is shown for different values of $g_0$. The coupling is calculated as before, using only the mode profile at the central frequency $M_i(x,y)$, and using the frequency dependent mode profiles $M_i(x,y,\omega)$. In every case the coupling is the same at the central frequency (1829\,rad/ps), but it differs outside of that. We use the global spectrum $\exp{(-(\delta\omega/\Delta\omega)^2)}$ as a point of comparison. With $g_0=-1$ the coupling to mode 6 will agree exactly with the global spectrum only when not considering the frequency-dependent mode profiles. When $g_0=0$, the coupling agrees with the global spectrum when including frequency dependence of the mode profiles---the mode size dependence of $M_6(x,y,\omega)$ agrees with the beam size $\propto1/\sqrt{\omega}$. When $g_0=1$ they once again differ, but in a different manner than with $g_0=-1$. In all three cases the true mode coupling calculated with $M_i(x,y,\omega)$ differs from the less complete calculation with $M_i(x,y)$, meaning that this form of amplitude STC caused by the consideration of $g_0$ matters. As a point of reference, the measurement of few-cycle laser systems has shown a $g_0$ of -2~\cite{hoff17-1}, but this will strongly depend on the system architecture, which is why we use -1, 0, and +1 as examples.

With zero offset, modeling with the frequency-dependent modes matters only for the amplitude, but with offset in $z$ it will also affect the spectral phase. This effect is significantly smaller for mode 1, and significantly larger for mode 15, and would also influence the coupling to other modes when either the offset is non-zero or there are other STCs that cause a break in cylindrical symmetry. Importantly, this effect is much more significant in the GRIN fiber compared to the step-index fiber. This is because in the GRIN fiber the mode size varies more strongly with wavelength.

\subsection{Other types of multi-mode waveguides}
\label{sec:rect}

As mentioned earlier, there are of course many different types of waveguides that have multiple spatial modes. One such case is gas-filled hollow-core fibers, a mature platform for pulse spectral broadening and nonlinear optics. Although they have a step-index profile and thus the losses are higher in the higher-order modes, the diameters of hollow-core fibers are so large, approaching 1\,mm, such that the scaling is favorable. Multi-mode solitons~\cite{safaei20}, visible pulse generation~\cite{piccoli21}, and UV dispersive wave generation and supercontinua~\cite{brahms21,grigorova23} have all been observed in hollow-core fibers thanks to significant participation of the higher-order modes. Besides the step-index and GRIN fibers already shown, multi-core or photonic-crystal fibers will also have interesting multi-mode solutions and dynamics~\cite{tani14,dupiol18,tishchenko22}.

\begin{figure*}[htb]
	\centering
	\includegraphics[width=\linewidth]{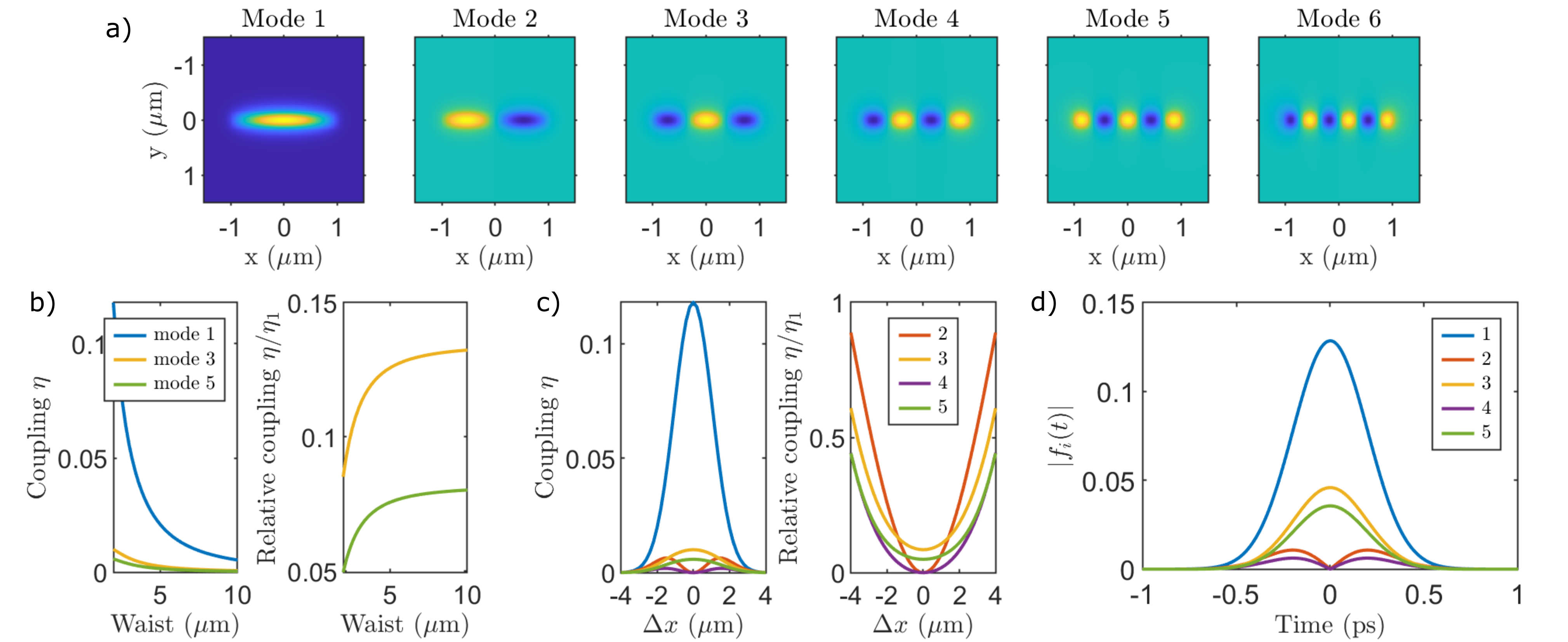}
	\caption{Example of coupling to a rectangular waveguide. The guided modes (a) at 1550\,nm wavelength for a ridge waveguide of Silicon on Silicon-Oxide with a height of 250\,nm and a width of 2\,$\mu$m. The coupling as a function of Gaussian beam waist (b) both absolute (left) and relative to the first mode (right). The coupling as a function of Gaussian beam offset (c) both absolute (left) and relative to the first mode (right) when the waist is 4\,$\mu$m. The amplitude of the coupling in time (d) when the input beam has a spatial chirp $\tau_t=\tau_0$ for a pulse with $\tau_0=200$\,fs and a 4\,$\mu$m waist.}
	\label{fig:rect}
\end{figure*}

Beyond optical fibers, integrated waveguides that are generally rectangular and smaller cross-section can still support multiple modes due to their large index contrast. The higher-order modes have been utilized for supercontinuum generation in straight waveguides~\cite{chenH21} and nonlinear optics in microresonators~\cite{zhaoY20,nitiss22,huJ22}, along with other parametric processes~\cite{signorini18,lacava19,franz21}. Due to the high level of confinement of the field the nonlinear optical modelling of these systems is much more complicated, where longitudinal fields can become important~\cite{driscoll09,poulvellarie20,ciret20}. Bridging the gap between CW multi-mode integrated photonics~\cite{daiD12,liC19,sunC20} and ultrafast space-time beams is a very exciting prospect.

Due to their small size, integrated waveguides will support a smaller total number of modes. Such integrated waveguides also usually have a step-index profile (limited by surface roughness and manufacturing accuracy), which leads to the same issues as with step-index fibers that the losses of the higher-order modes are significantly higher. This limit somewhat the amount of higher-order mode participation in nonlinear optical processes. For these reasons, among others, the total range of multi-mode phenomena has been more limited than with optical fibers. An interesting potential avenue is integrated semiconductor waveguides that have a graded-index profile~\cite{ocier20,porte21}, which may enable richer multi-mode physics.

To take a step back to the linear coupling with STCs that we have calculated earlier with the GRIN fiber, the same can be done for any other waveguide as long as the mode profiles are known. Intuition and symmetry considerations can predict the qualitative behavior. An example of various couplings can be shown for a rectangular waveguide in Fig.~\ref{fig:rect}(a). The 6 guided scalar modes are shown at 1550\,nm wavelength for the ridge waveguide of Silicon on Silicon-Oxide with a height of 250\,nm and a width of 2\,$\mu$m. The coupling to this waveguide also depends on the waist and offset, as shown in Fig.~\ref{fig:rect}(b--c). The key difference is that since the waveguide is so small, even with tight focusing there is a significant total loss when in-coupling---a significant portion of the energy does not couple to these propagating modes. Excitation of modes 3 and 5 (modes that are even in $x$) can be relatively quite large compared to the lowest-order mode, especially when the waist is larger. Adding an offset in $x$ allows one to couple to the odd modes as well, shown in Fig.~\ref{fig:rect}(c) for a 4\,$\mu$m waist. Such simple offset has already been used to tune supercontinuum in nanowaveguides~\cite{kou23}.

Just as with the GRIN fiber, STCs will create a frequency-dependent complex coupling $\mu_i(\omega)$ that represents the complex amplitude of the field coupled to each mode. An example with spatial chirp is shown when coupling to the rectangular waveguide in Fig.~\ref{fig:rect}(d). Because the waveguide is so small, the coupling to modes 2 and 4 is less than expected, since the 4\,$\mu$m waist is still relatively large---with a smaller waist there would be less coupling to modes 3 and 5 at $\omega_0$ but more coupling to modes 2 and 4 at other frequencies due to the spatial chirp.

\section{Conclusion}
\label{sec:conclusion}

We have outlined the linear coupling of free-space pulses having space-time couplings to multi-mode waveguides. We specifically considered in detail a graded-index (GRIN) fiber that supported many modes, but also a rectangular semiconductor ridge waveguide. The space-time couplings on the input pulse result in a different complex amplitude coupled into each mode: the spectral phase and amplitude depend on mode number, i.e. each mode has a different temporal field. The symmetries of the waveguide considered and the specific space-time coupling help to determine which modes are coupled to and why they have a specific spectral amplitude or phase.

As mentioned in the introduction, we have not discussed the propagation in the waveguide after the coupling takes place. In the linear case this is rather straightforward, where the group-velocity and group-delay dispersion of each mode determine how the modes will walk-off and spread in time, respectively. Only very recently have space-time effects at the input of a fiber been considered to help explain nonlinear optical processes observed in experiments, where the space-time propagation was modeled in the gas before entering into a hollow-core fiber~\cite{grigorova23}. We hope in the future to see how such space-time initial conditions will effect more exotic nonlinear processes in multi-mode fibers. One application may be to optimize multi-mode supercontinuum generation in a similar manner that has been done in the 1D case in the past~\cite{wetzel18}.

We discussed how the combination of simple space-time couplings could target a certain desired coupling, and how this may be extended to arbitrary space-time shaping on the input. However, since this is such an unexplored topic there are not known applications nor known targeted inputs. We hope that in parallel to the ongoing development in coupling to multi-mode waveguides, there will be study of the optimal multi-mode input fields so that the design and control process can be richer. Additionally, adding the additional degrees of freedom of polarization of orbital angular momentum, and their frequency or spatial dependence, are an exciting prospect for even more complete control of the initial field and therefore more complete control of resulting nonlinear optical physics.

\section*{Acknowledgements}

S.W.J. has received funding from the European Union’s Horizon 2020 research and innovation programme under the Marie Skłodowska-Curie grant agreement No 801505.

\bibliographystyle{unsrt}
\bibliography{biblo}
\end{document}